# The Arctic surface climate in CMIP6: status and developments since CMIP5

Richard Davy[1,2*] and Stephen Outten[1,2]
[1]Nansen Environmental and Remote Sensing Center, Bergen, Norway
[2]Bjerknes Center for Climate Research, Bergen, Norway

Here we evaluate the sea ice, surface air temperature, and sea-level-pressure from 31 of the models used in the Coupled Model Intercomparison Project Phase 6 (CMIP6) for their biases, trends, and variability, and compare them to the CMIP5 ensemble and the ERA5 reanalysis for the period 1979 to 2004. The principal purpose of this assessment is to provide an overview of the ability of the CMIP6 ensemble to represent the Arctic climate, and to see how this has changed since the last Phase of CMIP. Overall, we find a distinct improvement in the representation of the sea ice volume, but also in the sea ice extent, mostly linked to improvements in the seasonal cycle in the Barents Sea. However, numerous model biases have persisted into CMIP6 including too-cold conditions in the winter (4 K cold bias) and a negative trend in the day-to-day variability over ice in winter. We find that under the low emission scenario, SSP126, the Arctic climate is projected to stabilize by 2060 with a sea ice extent of around 2.5 million $km^2$ and a temperature 4.7 K warmer than the early $20_{th}$ century average, compared to 1.7 K of warming globally.

## 1. Introduction

The Arctic is of special importance in the study of Earth's climate system as it is especially sensitive to changes in global forcing, such as the enhanced forcing from the build-up of greenhouse gases (GHGs). This is exemplified by Arctic amplification: during the latter half of the twentieth century the Arctic has warmed at around twice the rate of the global average temperature, and this is most pronounced in the winter (Lu and Cai, 2009). There are numerous factors behind this Arctic amplification. While much attention has been given to the sea-ice albedo feedback (Serreze et al., 2009; Kumar et al., 2010), the recent Arctic amplification has been shown to be predominantly driven by changes in the outgoing longwave radiation (OLR) and not changes in the albedo, which primarily affects the absorption of shortwave radiation (Winton, 2006; Graverson and Wang, 2009). There are numerous processes which affect the OLR such as the Planck feedback (Planck, 1901), the lapse-rate feedback (Manabe and Wetherald, 1975), the water-vapour feedback (Graverson and Wang, 2009), changes in the atmospheric (Overland and Wang, 2010) and oceanic (Spielhagen et al., 2011) heat transport, and changes to the cloud cover (Vavrus, 2004). These changes to the surface energy budget lead to strong temperature changes in the Arctic due to the persistent stable stratification found in this region (Davy and Esau, 2016). This signal of Arctic amplification is robust and has also been identified in paleo-climate records (Dahl-Jensen et al., 1998; Masson-Delmotte et al., 2006; Brigham-Grette et al., 2013). Global climate models need to be able to capture this important feature of climate change so they must include a reliable representation of the relevant processes. A recent review of the relative importance of these processes in contributing to Arctic amplification within the Coupled Model Intercomparison Project (CMIP) phase 5 (CMIP5) global climate model results indicated that it is local temperature feedbacks which are largely responsible for the recent Arctic amplification (Pithan and Mauritsen, 2014). This is the process whereby the warmed air is trapped near the surface by the persistent stable stratification found in the Arctic, which leads to a greater warming in the Arctic than elsewhere (Esau et al., 2012).

Understanding the Arctic climate processes, and being able to simulate them within a global climate model, is essential if we are to understand the future climate in the Arctic as we go towards a new climatology of a 'Blue Arctic' i.e. nearly ice-free summers. This is a dramatic shift in the Arctic climatology and will bring profound changes to the natural environment in the region (Descamps et al., 2017), as well as the potential for human activities in the region e.g. through the opening up of shipping lanes (Smith and Stephenson, 2013; Melia et al., 2016). As such there has been a lot of focus on when the Arctic will become (nearly) ice-free in the summers (Overland and Wang, 2013). Attempts to estimate when this will happen have been done either by extrapolating from the observational record of sea-ice volume (Schweiger et al., 2011; Maslowski et al., 2012) or through analysis of global climate model projections (Pavlova et al., 2011; Stroeve et al., 2012; Wang and Overland, 2012; Massonnet et al., 2012). Sea-ice volume is chosen as the metric of interest for extrapolation from observation (rather than sea-ice area) since this has been shown to provide a more comprehensive measure of the evolution of Arctic sea-ice as it has shown a more rapid decrease than sea-ice area during the satellite era (Kwok et al., 2009; Stroeve et al., 2012). Indeed, in the previous Phase of CMIP it was demonstrated that sea-ice thickness was the primary cause of uncertainty in the sea-ice evolution during the $20_{th}$ and early $21_{st}$ centuries (Boe et al., 2010). Since sea-ice is not prescribed in the historical scenario of CMIP there exist large differences between the models as to the sea-ice extent, volume, and variability. This is especially important for projections of sea-ice decline as it has been demonstrated that the rate of loss of sea-ice depends upon the amount of sea-ice (Massonnet et al., 2012). Hence the CMIP model ensemble is often sub-sampled based upon the performance of individual models in describing the climatology and evolution of sea-ice during the satellite era (1979-present). This has been shown to be an effective and reliable method for constraining the large uncertainty that stems from the CMIP5 ensemble as to when we can expect ice-free summers in the Arctic (Massonnet et al., 2012).

Many small-scale processes have been identified as important in determining the Arctic climatology. These processes can be hard to represent accurately in global climate models, either due to the relatively coarse resolution of these models, or due to a limited understanding of these processes and their interactions. As such, the parameterization of small-scale processes can introduce biases into the Arctic climatology in these models. For example, Davy and Esau (2016) demonstrated the importance of shallow boundary layers in enhancing climate forcing signals, which can be very important in the Arctic which frequently has shallow, stably stratified boundary layers. However, global climate models have systematic biases towards over-estimating boundary-layer mixing under stable stratification (Seidel et al. 2012; Davy, 2018), which has been shown to lead to significant under-estimation of surface air temperature response to forcing (Davy and Esau, 2014). There are also systematic biases introduced due to the representation of mixed-phase clouds (Pithan et al., 2014; Tan et al., 2016), sea-ice albedo (Karlsson and Svensson, 2013; Koenigk et al., 2014), sea-ice extent (Stroeve et al., 2012), sea-ice variability and timing of the melting/freezing over the annual cycle (Mortin et al., 2014).

In section 2 we present the data and methods used in the paper; in section 3 we review the state of the Arctic climate in CMIP5 and CMIP6, and compare this to that in the ERA5 reanalysis; in section 4 we present the projections for the $21_{st}$ century under different forcing scenarios prescribed in CMIP6; and in section 5 we present our conclusions about the skill of CMIP6 models in capturing the current Arctic climate, the uncertainty in projections for the $21_{st}$ century, and how this picture has changed since the CMIP5 generation.

## 2. Data and methods

Here we use data from the CMIP5 and CMIP6 which have been made publicly available through the Earth System Grid Foundation web portal (https://esgf-data.dkrz.de/). For the CMIP5 simulations we use data from the historical and Representative Concentration Pathway (RCP) scenarios. And for the CMIP6 simulations we use data from the historical and Shared Socioeconomic Pathways (SSP) scenarios. For each of these scenarios we acquired the sea ice concentration and volume data at monthly resolution; the sea level pressure data at 6 hourly resolution; and the surface air temperature at daily resolution. The total Arctic sea ice extent and volume were calculated by multiplying the grid cell area by the sea ice concentration and thickness fields respectively, and then taking the sum of these for the whole northern hemisphere. For the other variables we applied a filter for the Arctic which selected only those data north of 66°N. The full list of CMIP5 and CMIP6 models used are presented in Table 1 and 2, along with the availability of the variables we have used in the different scenarios. There are quite large differences in the number of models for which a given variable is available for a given scenario. For example, there are many CMIP6 models which have surface air temperature and sea ice concentration data available, but not sub-diurnal sea level pressure and neither monthly sea ice volume. We compare the CMIP model results to the European Centre for Medium Range Weather Forecasting's ERA5 reanalysis for the surface air temperature, sea level pressure, and sea ice extent; and we compare the sea ice volume from the climate models to that from the PIOMAS reanalysis (Zhang and Rothrock, 2003). We chose to compare the climatology from the models and the reanalysis over the period 1979-2004 because this covers the period of satellite observations (from 1979) when the reanalysis is well-constrained by observations and finishes at the end of the historical scenario protocol of the CMIP5 in 2004. We therefore argue that choosing this period provides the fairest comparison between the CMIP5 and CMIP6 model results.

For each monthly timeseries we converted the data to anomalies by removing the full-period climatological mean for each month. For the daily surface air temperature data, we took an area-weighted mean over the Arctic, and then calculated the standard deviation within each month to create a monthly timeseries describing the day-to-day variability in the Arctic. We also took the standard deviation within each month of the daily-mean temperature for each gridpoint to create a monthly timeseries for each location. This process was repeated for the 6 hourly sea level pressure data to create a timeseries of the intra-monthly variability in the Arctic-mean sea level pressure and a monthly timeseries for each grid cell of the intra-monthly variability in sea level pressure. We then converted each of these variables into anomalies by removing the climatological average for each month, calculated from the full period.

The intra-annual auto-regression was calculated by taking a linear regression of the anomaly in a given month against the anomaly from the previous month. For example, the intra-annual auto-regression of temperature in March was calculated by taking the linear regression of the March temperature anomalies against the February temperature anomalies. This process was repeated for all variables for which we calculated the intra-monthly auto-regression.

## 3. Representation of the present climate

### (A) Sea-ice

#### Climatology

Sea ice cover is crucial in determining the climatology of the Arctic region. This is because the presence of sea ice acts to decouple the exchange of heat, moisture, momentum, and particles between the ocean and atmosphere. Therefore, differences in the sea ice concentration and extent can lead to large differences in the surface energy budget, and consequentially the Arctic climatology. Sea ice thickness has been noted as one of the largest sources of uncertainty in the evolution of sea ice (Zygmuntowska et al., 2014). There is an inverse relationship between sea ice thickness in the Arctic basin and sea ice export, both in the inter-annual variability and in the long-term trends, that has been demonstrated to hold for a selection of CMIP5 models (Langehaug et al., 2013).

Figure 1 shows the climatology of the total Arctic sea ice extent and volume in CMIP5, CMIP6, and the ERA5 reanalysis for the period 1979-2004. The sea ice extent in the ERA5 reanalysis varies from around 17.4 million km$_2$ at the peak extent in March to a minimum of around 7.3 million km$_2$ in September. Both the CMIP5 and CMIP6 model ensembles capture the seasonal cycle with a close agreement between the multi-model medians and the ERA5 climatology – at no point in the annual cycle does either multi-model median have a bias of more than 1 million km$_2$. Overall the CMIP6 ensemble median has a closer agreement to the ERA5 than does the CMIP5 median due to a better fit in the seasonal minima in September. However, both the CMIP5 and CMIP6 medians underestimate the seasonal maximum extent by almost 1 million km$_2$. Individual models in both the CMIP5 and CMIP6 ensembles can have very large biases in the sea ice extent of up to 7 million km$_2$, but model biases are still small compared to the amplitude of the seasonal cycle.

The climatology of the sea ice volume presents a different picture. In PIOMAS the Arctic sea ice volume changes from a peak of almost 30,000 km$_3$ in April, to a minimum of around 13,400 km$_3$ in September. The multi-model medians of both the CMIP5 and CMIP6 ensembles have a similar climatology to that of PIOMAS although they both underestimate the sea ice volume at the seasonal minima and so overestimate the amplitude of the seasonal cycle (Figure 1). There is a slightly reduced bias in the CMIP6 ensemble median compared to CMIP5 due to a closer fit at the seasonal maxima, but the difference between the CMIP5 and CMIP6 medians is small. However, there is a very large in both ensembles, bigger than the amplitude of the seasonal cycle.

While there is some indication of improvements in capturing the climatology of both the sea ice extent and volume with the ensemble median of CMIP6 compared to CMIP5, the large inter-model spread in the sea ice volume demonstrates there is still very large uncertainty in historical sea ice volume in these models.

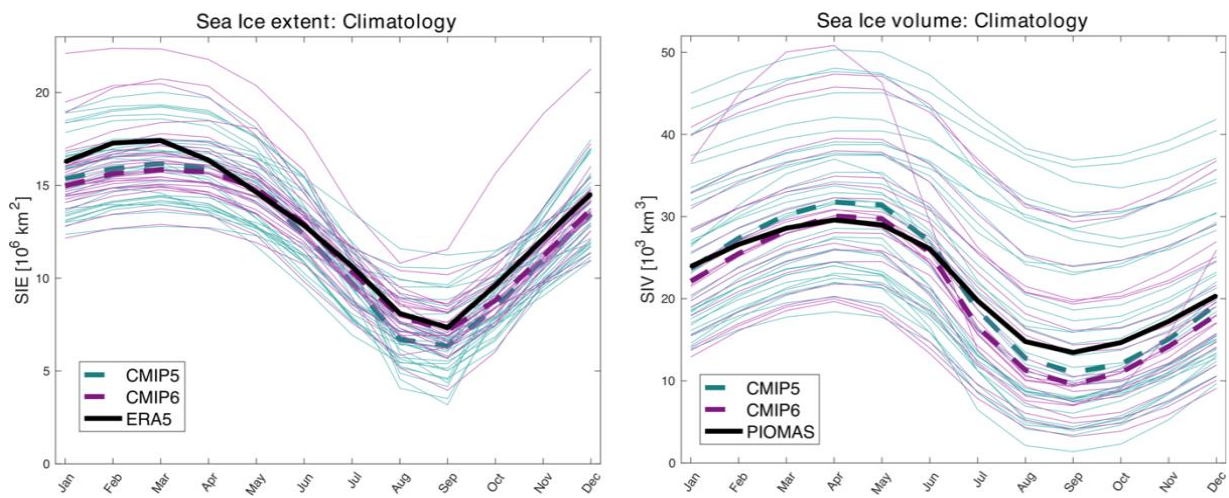

*Figure 1. The climatology of the Arctic sea ice extent (left) and volume (right) for the period 1979-2004. The CMIP5 models (turquoise) and the CMIP6 models (purple) are shown along with their multi-model means (thick dashed line). The thick black line shows the climatology from the ERA5 for the extent, and PIOMAS for the volume.*

Figure 2 shows the sea ice thickness for the CMIP5 and CMIP6 ensemble median, and the difference between the two ensembles, for the months of March and September and for the annual mean. The climatological mean sea ice thickness has a remarkable geographical consistency between the CMIP5 and CMIP6 ensembles in all months and in the annual mean. In the annual mean the CMIP6 ensemble mean has much thicker (up to 1 m) sea ice in the Canadian archipelago and somewhat thicker (≈ 30 cm) ice in the Fram strait compared to the CMIP5 ensemble. There is also thick ice in many of the Canadian lakes in the CMIP6 ensemble which was not present in CMIP5. The pattern of thicker ice in CMIP6 over the Canadian archipelago can be found in all months. The Canadian archipelago is a particularly challenging region for sea ice modelling in climate models because of the highly broken land cover and the challenges of capturing sea ice interactions with the land in such complex terrain (Kwok, 2015). The simulated climatology of sea ice in this region is therefore sensitive to both the horizontal resolution of the model and the sea ice physics.

At the seasonal minima in March we see the CMIP6 ensemble has thinner ice over most of the Arctic which is reflected in the lower overall sea ice volume seen in Figure 1. The biggest difference in the September climatology is in the northernmost Canadian archipelago: in CMIP6 the models have relatively thick ice here (> 2 m), whereas in CMIP5 the ice in this region was relatively thin or not present. In the CMIP6 ensemble the ice is also generally thicker (20 to 50 cm) across most of the Arctic basin. In the annual-mean these seasonal differences in the Arctic basin largely cancel out, and the difference between the CMIP5 and CMIP6 ensembles is small compared to the mean thickness (< 10 cm across most of the Arctic).

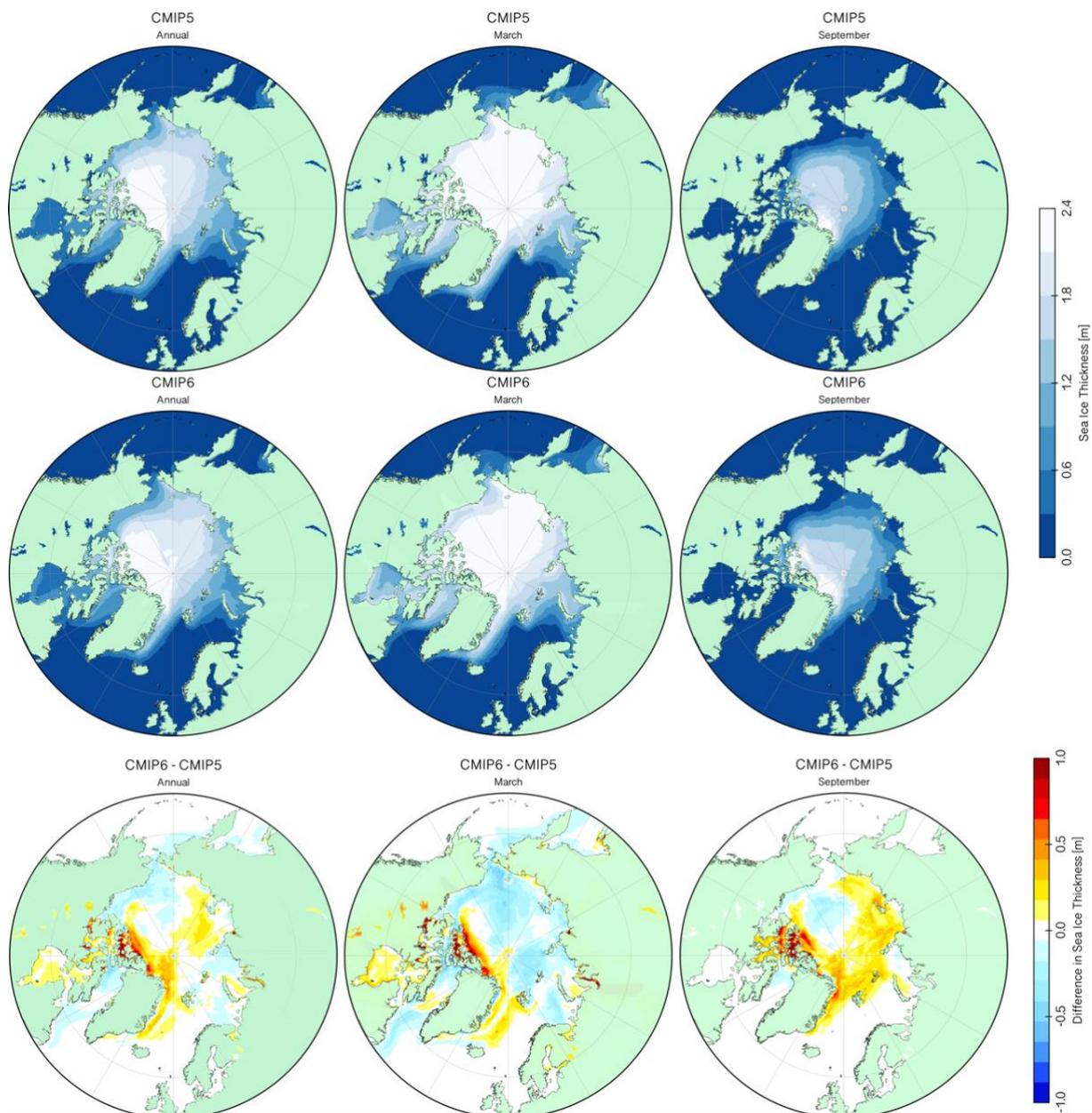

*Figure 2. The climatological mean sea ice thickness from the CMIP5 and CMIP6 ensemble median, and the difference between the two ensembles, for the annual-mean and for the months of March and September, for the period 1979-2004.*

Figure 3 shows the bias in the sea ice thickness for the CMIP5 and CMIP6 ensemble medians with respect to the PIOMAS simulations for the months of March and September and the annual mean. The sea ice extent from the ensemble-median and the ERA5 reanalysis are also highlighted with thick red and black lines respectively.

First, we can see that in almost all locations and times both the CMIP5 and CMIP6 ensemble medians have thinner sea ice than is found in PIOMAS. This is consistent with our expectation from Figure 1 where we saw that the ensemble medians have lower sea ice volume in the Arctic than does PIOMAS. However, differences in sea ice extent and concentration can also affect the bias in sea ice volume. The bias in the sea ice extent is shown by the difference between the thick red lines and the black lines in Figure 3. We already saw that the CMIP6 ensemble has a better agreement to the observed sea ice extent than does the CMIP5 ensemble (Figure 1), and this is reflected in the annual-average extent in these two ensembles (Figure 3). Much of the improvement in the CMIP6 ensemble comes from a better fit to the observed extent in the Barents Sea. We can see from the sea ice extent in March that this is due to an improved representation of the Barents Sea extent during the seasonal minima. Despite this improvement the CMIP6 models are still biased towards having too much sea ice in the Barents Sea in winter. In the CMIP5 ensemble the models tended to over-estimate the sea ice extent in winter which introduced large biases into the climate of the region and led to an intense focus on the processes of sea ice removal and formation in this region (Smedsrud et al., 2013).

There is generally a good fit to the observed sea ice extent extent in all other regions, except for a too-high extent in the Fram strait which is persistent in both the CMIP5 and CMIP6 ensembles. The climatologies in the seasonal maxima (March) and minima (September) in sea ice extent more clearly show the seasonality of this bias. In March both the CMIP5 and CMIP6 ensembles have a much too high extent in the Fram strait with sea ice extending to the coast of Iceland, which has not been seen in the ERA5 reanalysis. This is likely linked to biases in the winter sea ice export through the Fram strait (Langehaug et al., 2013). In contrast, both the CMIP5 and CMIP6 ensembles have a good fit to the observed extent in the Fram strait in September, with the CMIP6 ensemble having an excellent fit to the observations at this time of year. However, another bias in September sea ice extent which has persisted between the CMIP5 and CMIP6 ensembles is that both have too little sea ice in the Kara sea at the seasonal minima. This is a region where the September sea ice has been retreating during the period 1979-2004.

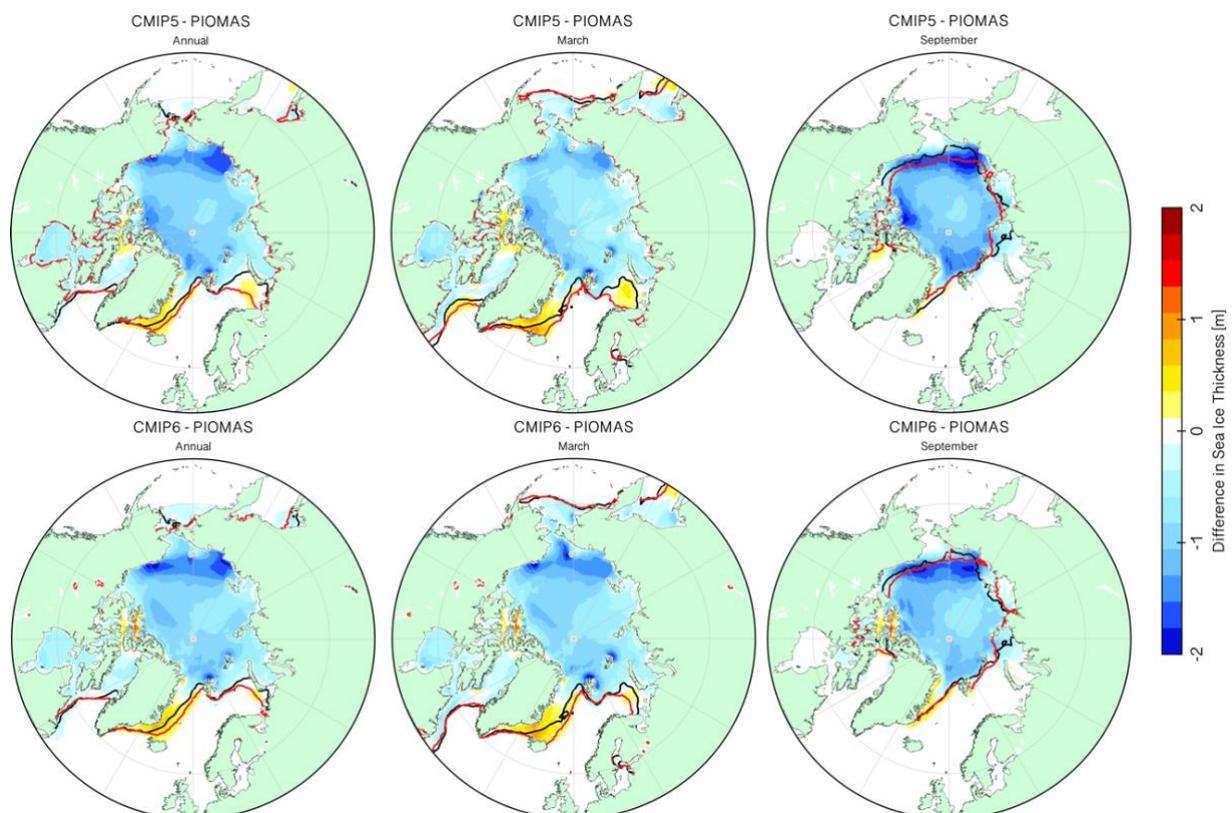

*Figure 3. The difference in the climatological mean sea ice thickness from the CMIP5 and CMIP6 ensemble mean and the PIOMAS simulations for the months of March and September, and for the annual-mean for the period 1979-2004. The climatological mean sea ice extent from ERA5 is shown by the thick black line and from the ensemble mean by the thick red line.*

## Trends and variability in the historical period

This climatology is taken during a period of rapid decline in both the Arctic sea ice extent and volume. It is therefore necessary to also evaluate the current ensemble of climate model's ability to capture the rate of decline in Arctic sea ice over this period. Figure 4 shows the trend in the sea ice extent and volume from the CMIP5 and CMIP6 ensembles compared to those from ERA5 and PIOMAS respectively. Almost all the CMIP model results for the historical period show a decline in the sea-ice extent and volume during the period of our evaluation, 1979-2004. However, there is a very large spread in the trends between the different models. No individual model in either the CMIP5 or CMIP6 ensembles has a better fit (as measured by mean absolute difference) to the trends in sea ice extent from ERA5 than do the multi-model means.

In ERA5 there is a negative trend in the sea ice extent in all months, but there is also a pronounced seasonal cycle in the trend in sea ice extent with the most rapid decline in the summer months of July-August-September. The slowest decline occurs in December when the trend is -25,000 $km^2$ $yr^{-1}$ and the fastest decline occurs in September when the trend is -55,000 $km^2$ $yr^{-1}$, although this is very similar to the trends in July and August. This seasonal cycle is to some extent captured in the CMIP5 and CMIP6 multi-model means which both have the peak decline in September. The CMIP6 ensemble mean has a better fit to the observed trends in September, and in the annual average, than does the CMIP5 ensemble mean. However, there is a very large spread in both the CMIP5 and CMIP6 ensembles with many models not having a clear seasonal cycle.

The trend in sea ice volume has much less of a seasonal cycle than the trend in extent, as might be expected. In PIOMAS the trend is very similar in all months with an average decrease of -204 $km^3$ $yr^{-1}$. There is a weak seasonal cycle in PIOMAS with a minimum trend of -189 $km^3$ $yr^{-1}$ in February and a peak trend of -223 $km^3$ $yr^{-1}$ in July. Due to the small seasonal cycle, model biases in the trend in sea ice volume are principally systematic biases rather than seasonally dependent. Both the CMIP5 and CMIP6 models have a very large spread in the trends in sea ice volume. The CMIP6 ensemble mean is a better fit to the trends found in PIOMAS than is the CMIP5 ensemble mean, but there is an extremely large spread in the ensemble indicating that the response of the sea ice volume over the historical period is still very poorly captured by the CMIP6 models.

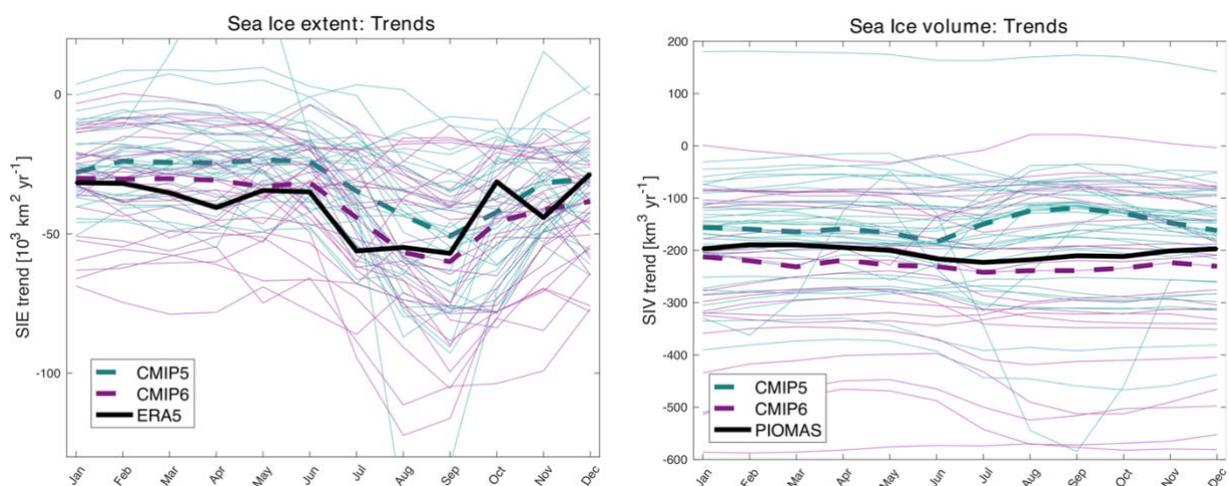

*Figure 4. The trend in the Arctic sea ice extent (left) and volume (right) for each month in the period 1979-2004. The CMIP5 models (turquoise) and the CMIP6 models (purple) are shown along with their multi-model means (thick dashed line). The thick black line shows the climatology from the ERA5 for the extent, and PIOMAS for the volume.*

Another measure of the skill of the CMIP models in capturing the physics of sea ice, and sea ice driving forces, is to determine the degree of red noise in the system i.e. how much the anomalies in one month are related to those in the previous month. Figure 5 shows the 1-month autoregressions in anomalies of sea ice extent and volume for the CMIP5 and CMIP6 ensembles, in comparison to that found in ERA5 and PIOMAS for the extent and volume respectively. We can see that in the ERA5 reanalysis the sea ice extent has a generally high autoregression which ranges from R=0.54 in June to R=0.90 in September. However, this is almost always lower than the autoregression found in the CMIP5 and CMIP6 models. The model ensembles both have two distinctive peaks in the seasonal cycle, one in March at the seasonal maxima in extent, and one in September at the seasonal minima in extent. However, this seasonality of predictability of anomalies in extent is not found in the ERA5 reanalysis.

The autoregression of sea ice volume anomalies is even higher than that for extent. In the PIOMAS there is a clear seasonal cycle to the auto-regression where it ranges from a minimum of R=0.92 in June and July to a maximum of R=0.99 in December. As with the extent, the autoregression in the reanalysis is somewhat lower than that found in the models. In both the CMIP5 and CMIP6 ensemble means the autoregression is always higher than that of PIOMAS, ranging from R=0.97 to R=0.99. However, in both the CMIP5 and CMIP6 ensemble means there is a similar seasonal cycle as that found in PIOMAS with a minimum in the autoregression in July.

There is a clear bias in both the CMIP5 and CMIP6 model ensembles to having a too-high persistency of anomalies in both sea ice extent and volume within the seasonal cycle. The clearest example of that in the models is the predictability of September sea ice extent anomalies in the models based on the August extent: there is very little spread between the models, and they all have an extremely high predictability of the September sea ice extent. This could lead to over-confidence in the seasonal predictability of sea ice extent and volume from these models.

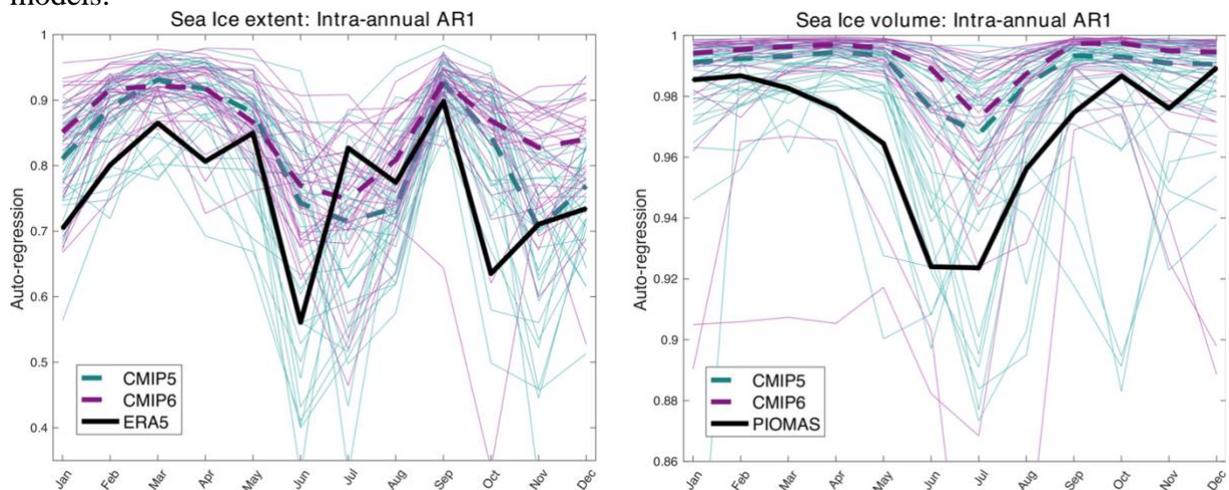

*Figure 5. The auto-regression of the Arctic sea ice extent (left) and volume (right) for each month in the period 1979-2004. The CMIP5 models (turquoise) and the CMIP6 models (purple) are shown along with their multi-model mean (thick dashed line). The thick black line shows the climatology from the ERA5 for the extent, and PIOMAS for the volume.*

## (B) Surface air temperature

Figure 6 shows the climatology and the inter-annual variability of the Arctic-mean surface air temperature for each month from the CMIP5 and CMIP6 ensembles, and from the ERA5 reanalysis, for the years 1979-2004. The climatological-mean seasonal cycle is almost identical in the CMIP5 and CMIP6 ensemble means. In the CMIP5 and CMIP6 ensemble means the temperature varies from a minimum of 245 K in January and February to a maximum of 275 K in July. There is a larger spread in the CMIP5 ensemble in the winter temperatures (December-January-February). The ensemble means of both CMIP5 and CMIP6 are colder than the reanalysis throughout the whole year, but the largest difference with the ERA5 reanalysis is in the winter: the ensemble means are more than 4 degrees colder than the reanalysis in January and February.

This challenge with capturing the wintertime mean temperatures also extends to the inter-annual variability in winter temperatures. Throughout the wintertime both the CMIP5 and CMIP6 model ensemble means have a higher inter-annual variability than the ERA5 reanalysis. The biases in both the climatological mean and inter-annual variability of the surface air temperature are likely related to the differences in the climatology of the sea ice detailed above. If a model has too-much sea ice it will likely be colder than the reanalysis and have higher variability because the air temperature over ice is colder and more sensitive to changes in forcing than over open ocean. This can be seen from a geographical analysis of the model biases.

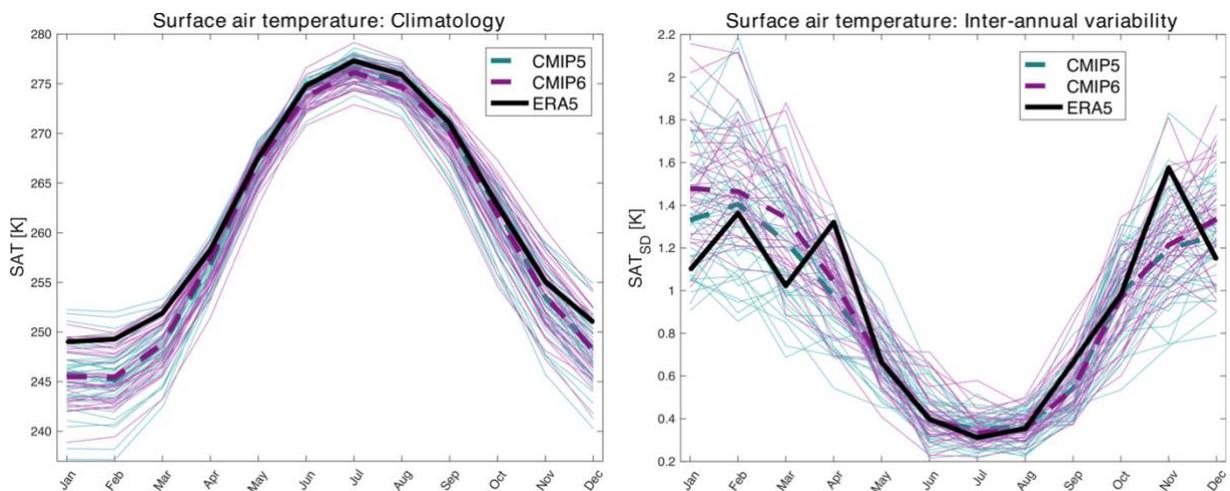

*Figure 6. The climatological mean and inter-annual variability of the surface air temperature averaged over the Arctic region for the years 1979 to 2004 from the CMIP5 ensemble (turquoise), the CMIP6 ensemble (purple) and the ERA5 reanalysis (black). The ensemble means are highlighted by the thick dashed lines.*

Figure 7 shows the map of the climatological mean surface air temperature for the CMIP6 ensemble mean, the difference between the CMIP5 and CMIP6 ensemble means for the for the annual mean, and for the months of January and July.

In the CMIP6 ensemble mean climatology the coldest region is of course over the interior of Greenland where the average surface air temperature is below the freezing point of water throughout the year. The largest differences between the CMIP6 and CMIP5 ensemble means are around the coast of Greenland and in the Barents Sea. In both of these regions the CMIP6 ensemble mean is warmer than the CMIP5 by around 2 K. The difference between these two ensembles is largest in the winter (January) in the Barents Sea, where temperature differences are around 4 K in the Western Barents Sea. This is directly related to the reduced sea ice cover in this region in winter (Figure 3) since without the insulating effect of sea ice, the surface air temperature is not able to cool. In the summer there is very little difference between the CMIP5 and CMIP6 ensemble means. There is almost zero difference over ice where the air temperature is limited by the melting of sea ice. But even over land the temperature difference between the two ensembles is small, with the CMIP6 ensemble generally being approximately 1 K colder over land than the CMIP5 ensemble.

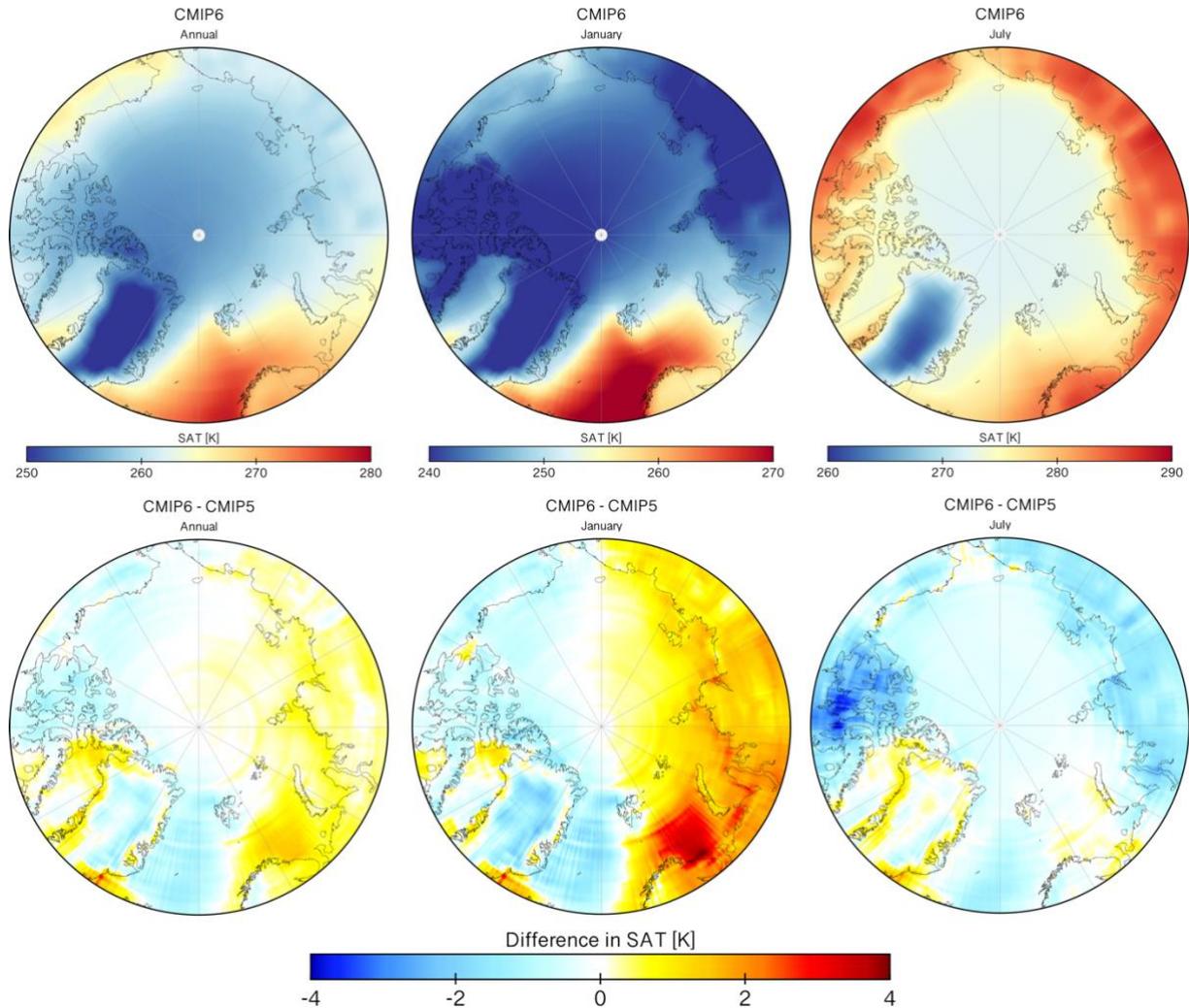

*Figure 7. The top three panels show the climatological mean surface air temperature from the CMIP6 ensemble mean and the lower three panels shows the difference between the climatologies in the CMIP6 and CMIP5 ensemble means for the annual mean and the months of January and July. The top panels are plotted on different scales to ensure the spatial details are visible.*

In comparison with the ERA5 reanalysis, we can see that both the CMIP6 and CMIP5 ensembles are more than 1 K colder than the reanalysis in the annual mean (Figure 8). Most of this bias is explained by differences in the wintertime temperatures as can be expected from Figure 6. In both the CMIP5 and CMIP6 ensemble means we find cold biases of up to 9K in regions with thick sea ice, such as off the coast of Northern Greenland, but also in the regions where there is sea ice in the models but not in the reanalysis, such as in the Fram Strait and in the Barents Sea (Figure 8). The bias in the CMIP6 ensemble over thick sea ice is very similar to what was found in CMIP5, and most of the bias reduction in CMIP6 is associated with an improved representation of the sea ice edge in the Barents Sea.

In the summer (July) the pattern of bias is almost identical in the CMIP5 and CMIP6 ensembles. There is a general cool bias across most of the Arctic with an almost 1 K cold bias over ocean, sea ice, and almost all the land. The only clear exceptions are over the interior of Greenland where the model ensembles can have up to 5 K warm biases compared to the reanalysis. There are also strong warm biases over coastal waters everywhere in the Arctic, but especially in the Canadian archipelago.

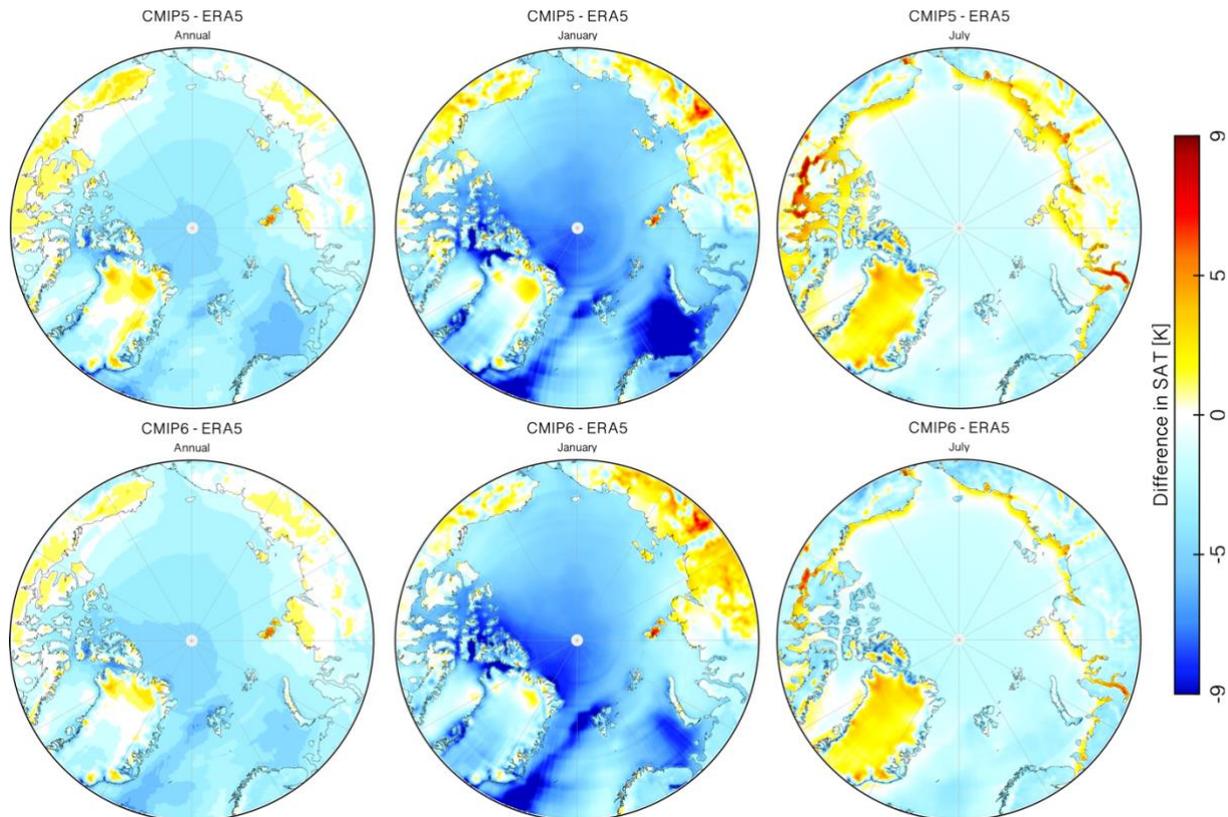

*Figure 8. The difference in the climatological mean 2m air temperature between the ERA5 reanalysis and the multi-model means of CMIP5 (top panels) and CMIP6 (lower panels) in the annual mean, and for the months of January and July over the period 1979-2004. The center panels for January are plotted on a different scale from the annual and July panels.*

A robust feature of the overall warming trend is the reduction of day to day variability in surface air temperature (Moberg et al., 2000). This signature of global warming is not unique to the Arctic, but it can be found in this region. One explanation for this reduction in the day-to-day variability in the Arctic is that since surface air temperatures over ice are more sensitive to forcing than the air over ocean (Esau et al., 2012) so the reduction in sea ice cover should also result in a reduction in the SAT variability. Another potential explanation for this is that as the surface warms the depth of the mixed layer increases which reduces the sensitivity of the surface air temperature to changes in forcing (Davy and Esau, 2016). We can differentiate between these two explanations by evaluating where the reduction in variability is occurring geographically and by relating the changes in variability to the changes in sea ice extent at the seasonal extrema. Figure 9 shows the trend in the day-to-day variability of surface air temperature from CMIP5, CMIP6, and the ERA5 reanalysis.

In the annual mean both the CMIP5 and CMIP6 ensemble means have similar trends: there is a negative trend almost everywhere in the Arctic, with the largest reduction in day to day variability over the Barents and Bering Seas. This pattern is very consistent between the two model ensembles but it is quite different to the pattern found in the reanalysis. In ERA5 there is a negative trend across much of the Arctic, but there is a strong positive trend in the area north of Greenland.

The difference in the patterns in the annual mean can be explained by looking at the trends at the seasonal extrema in sea ice extent in March and September. In March the model ensembles have a negative trend across most of the Arctic, aside from the CMIP5 ensemble which has a positive trend centered over the Canadian archipelago. However, this pattern is very different to what we find in ERA5 in March: at this time the reanalysis has a strong positive trend in the day-to-day variability across almost all the sea ice, and it is particularly strong in the area north of Greenland. This is the primary cause of the difference in the pattern of trends in the annual mean. During the seasonal minima in sea ice extent in September the model ensembles have better agreement with the reanalysis with a negative trend across all the sea ice, although in ERA5 there are very different patterns over land.

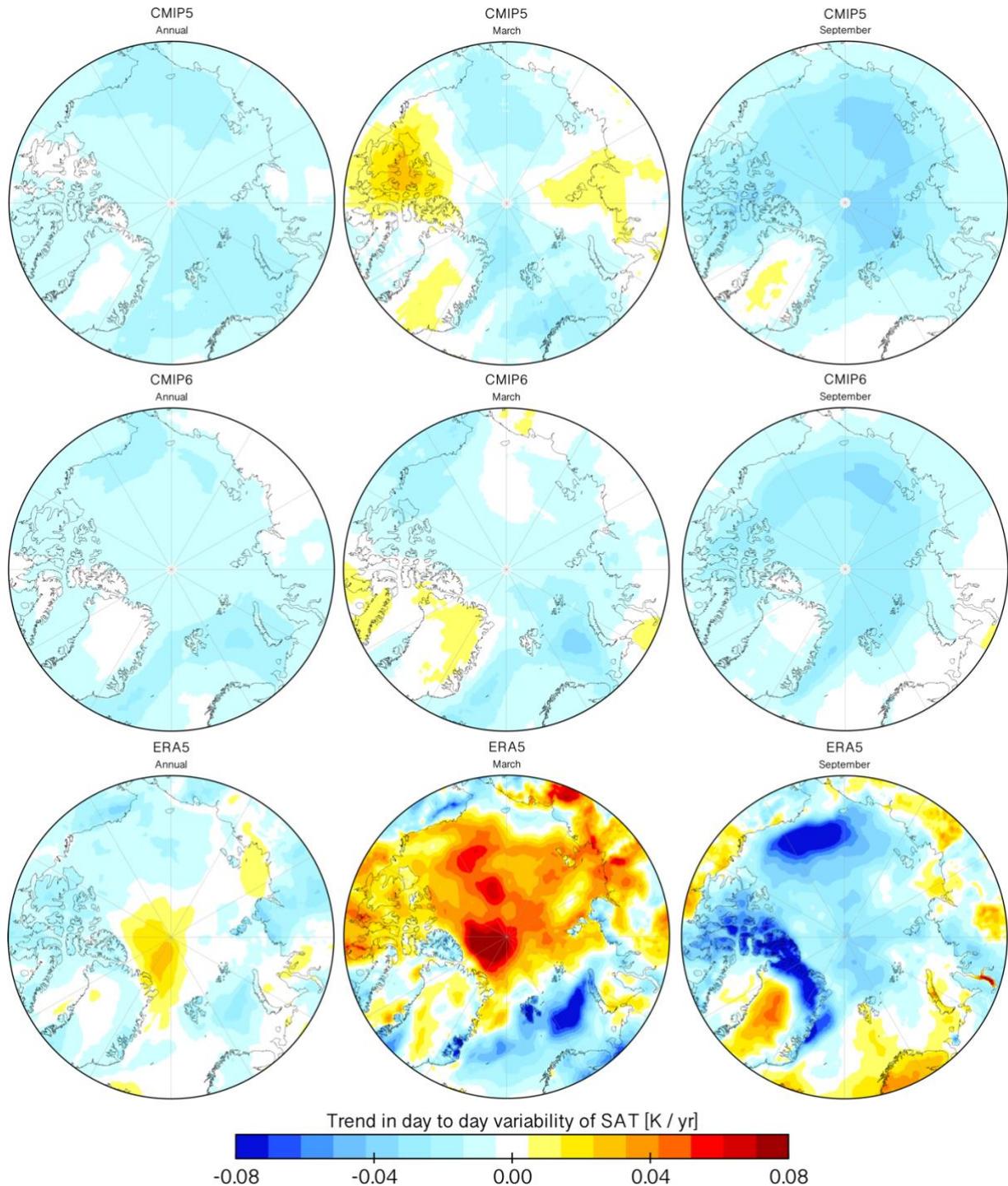

*Figure 9. The trend in the day-to-day variability in surface air temperature from the multi-model means of CMIP5 (top), CMIP6 (middle), and ERA5 (bottom) for the annual mean (left), March (middle), and September (right) over the period 1979-2004.*

This large discrepancy between the CMIP simulations and the ERA5 reanalysis in the day-to-day variability may have a number of causes, including biases in the synoptic activity (see next section). It may also be related to the biases in the climatology of the mean SAT (Figure 8). The models tend to be too cold in winter across the entire Arctic ocean, but especially in those regions with thick sea ice north of Greenland. Since colder conditions are more sensitive to changes in thermal or radiative forcing, this might explain why the models are biased towards too-high day-to-day variability in SAT, and consequentially do not have the same response to forcing. Another likely cause is the limitations of the sea ice physics and surface coupling in these climate models. Many processes such as the dynamics of sea ice or the exchanges over leads in the ice are either poorly captured or altogether missing in climate models. However, these may play an important role in determining the low-level stability and surface fluxes, thus affecting the day-to-day variability. But since the reduction in variability is not limited to those regions where sea ice has been retreating over the period 1979-2004, it is likely that this reduction in variability is not solely due to changes in sea ice extent.

### (C) Sea level pressure

Here we use the intra-monthly variability in sea level pressure as a measure of the atmospheric dynamics. The climatology and inter-annual variability of the intra-monthly variability in sea level pressure from the CMIP5 and CMIP6 ensembles, and the ERA5 reanalysis are shown in Figure 10. In the ERA5 reanalysis the intra-monthly standard deviation of 6-hourly sea level pressure reaches a peak of around 500 Pa in the winter months (January-February-March), to a minimum of around 220 Pa in July. Overall this is comparable to the multi-model means from CMIP5 and CMIP6. However, we can see from the CMIP5 and CMIP6 multi-model means that the climate models tend to over-estimate the intra-monthly variability, especially in the winter months. This tendency to over-estimate the amount of synoptic activity in the Arctic winter months may be related to the model biases in day-to-day variability, although there is no significant correlation between model biases in the SLP variability and the SAT variability in either the CMIP5 or CMIP6 ensembles.

The models also show too-high variability in the inter-annual variability of synoptic activity, especially in the winter months (Figure 10). Although the sudden drop in inter-annual variability in the January data in ERA5 is anomalous compared to the rest of the seasonal cycle form which we might expect a peak in this month. The CMIP5 and CMIP6 multi-model means are very similar in the inter-annual variability and have either similar or greater variability than is found in the ERA5 reanalysis in the same period.

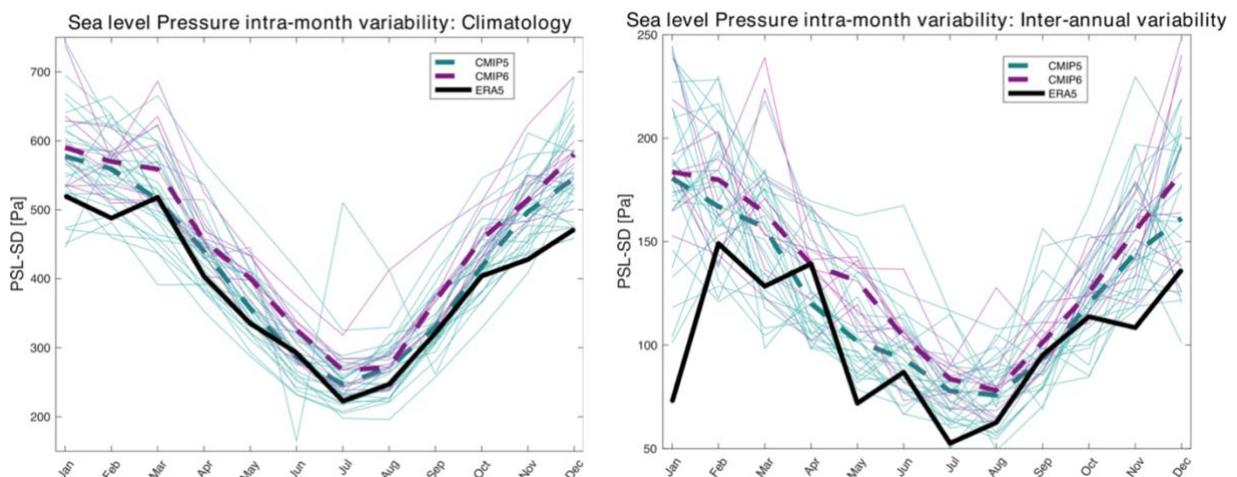

*Figure 10. The climatological mean and inter-annual variability of the intra-monthly standard deviation of the 6-hourly sea level pressure averaged over the Arctic region for the*

*years 1979 to 2004 from the CMIP5 ensemble (turqoise), the CMIP6 ensemble (purple) and the ERA5 reanalysis (black). The ensemble means are highlighted by the thick dashed lines.*

## 4. Projections for future change

The pace of climate change is currently fastest in the Arctic due to the various positive feedbacks found in this region (Serreze et al., 2007; Screen and Simmonds, 2010; Pithan and Mauritsen, 2012), and this is expected to continue throughout the 21$_{st}$ century (IPCC AR5, 2013). Even limiting the mean warming of the Earth to 1.5 degrees will still mean a warming of around 3 K in the Arctic combined with a large loss of sea ice (Niederdrenk and Notz, 2018). An important question under future forcing scenarios is how fast the Arctic sea ice will be removed and how confident we can be in those projections. The confidence can be assessed using the spread in the model ensemble running a given experiment, so it is highly relevant to assess model spread under the different CMIP6 forcing scenarios for the 21$_{st}$ century, and how this picture has changed since CMIP5.

### Sea ice extent and volume

Figure 11 shows the timeseries of sea ice extent and volume from the CMIP5 and CMIP6 ensembles with comparison to the ERA5 reanalysis for the sea ice extent and to PIOMAS for the sea ice volume over the period 1900-2100. The CMIP5 ensemble uses the RCP8.5 forcing scenario for the years after 2005, and the CMIP6 results use the SSP585 scenario after the year 2014. There is an almost identical sea ice extent in both the annual mean and the September minima in the CMIP5 and CMIP6 ensemble means prior to the 1990s. However, there is a very large spread in both the CMIP5 and CMIP6 ensembles during this period with almost all models having annual-mean extents in the range of 11-16 million km$_2$, while the September minima has an even larger spread in this period with almost all models in the range 5-12 million km$_2$. There is a very good fit between both the CMIP5 and CMIP6 ensemble medians and the observed sea ice extent in the annual mean and September.

After the 1990s the sea ice extent was observed to rapidly reduce for the next 20 years, especially in the seasonal minima in September (Figure 11). There was a slower reduction of sea ice extent in the CMIP5 ensemble than was observed since the mid-1990s. There has been a lot of speculation about the reasons for this (Stroeve et al., 2012). It could be because in the CMIP5 protocol in the post-2005 period only the CO2 forcing was changing in time, whereas other forcings might have been important. It is also likely that there are essential processes of sea ice coupling which are not included in these models which were important in causing this rapid decline. In the CMIP6 protocol the historical forcings extend up until 2014, and so cover part of this period of rapid decline (Eyring et al., 2016). There have also been efforts to address the depiction of sea ice in these models since the CMIP5 (Kattsov et al., 2010). Some combination of these changes has led to an improvement in the CMIP6 ensemble over that from CMIP5. We can see there is a more rapid decline in the sea ice extent over this period compared to CMIP5, and it is closer to the observed trend (Figure 4).

This pattern of a more negative trend in CMIP6 than in CMIP5 continues into the 21$_{st}$ century projections in comparison of the scenarios RCP8.5 and SSP585. These two 21$_{st}$ century scenarios are not directly comparable, but they both correspond to the upper range of possible rates of change for the 21$_{st}$ century in the CMIP5 and CMIP6 protocols respectively. In both the RCP8.5 and SSP585 scenarios we see a rapid decline in the sea ice extent, especially in September. Under the SSP585 scenario the multi-model mean reaches essentially ice-free conditions in September (less than 1 million km$_2$) by around 2060, while in the RCP8.5 this isn't expected to be reached until around 2080. The difference between the RCP8.5 and SSP585 scenarios is also large in the projections for the annual mean extent: in the multi-model mean of the SSP585 projections the Arctic only has around 4 million km2 by the end of the 21st century, whereas under RCP8.5 the CMIP5 models projected there to still be an annual average of around 6 million km2 of sea ice.

While there is some indication of improvement in CMIP6 in the representation of historical sea ice extent, there is a clear improvement in the multi-model mean of sea ice volume. In the period 1900-1980s when the effect of CO2 forcing was relatively weak, we can see there is still a very large spread in the CMIP6 ensemble projections of sea ice volume, comparable to that found in CMIP5. The multi-model means are also almost identical between CMIP5 and CMIP6 in this period. However, in the period from 1980 to present there was a very rapid decline in the sea ice volume, according to the PIOMAS reconstruction. While the CMIP5 underestimates the rate of decline of sea ice volume, this is extremely well captured by the multi model mean of the CMIP6 models, both in the annual mean and the September minima (Figure 11).

For the 21st century projections the annual-mean sea ice volume is projected to have a near-linear decrease in time under the SSP585 scenario, getting very close to zero to the end of the 21st century. This contrasts with the RCP8.5 projections where the models begin the simulations with a higher overall volume and decrease over the 21st century at a similar rate to that found in the SSP585 projections, thus ending the 21st Century with a volume of around 7000 km3 compared to just 1000 km3 in the CMIP6 ensemble.

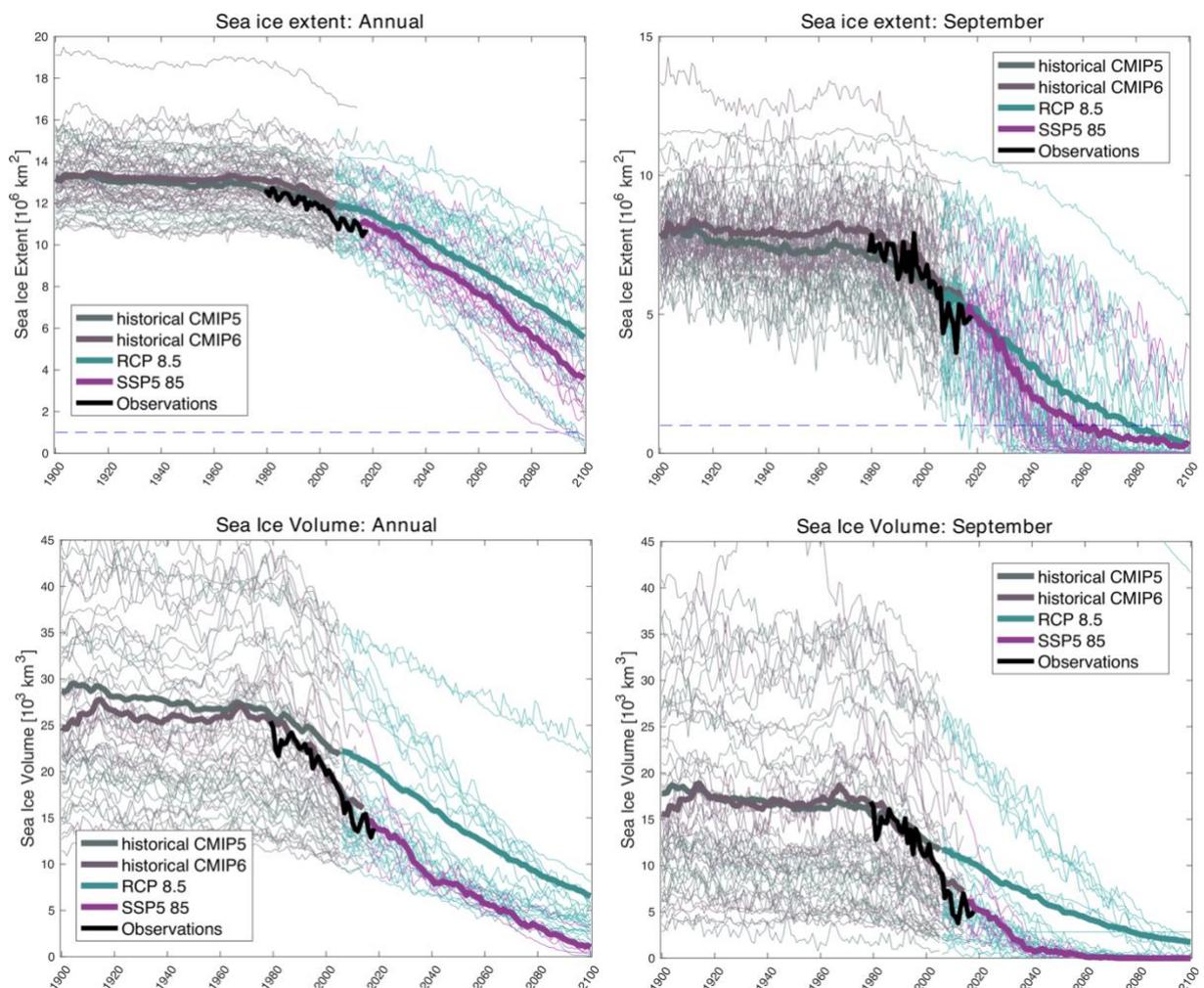

*Figure 11. Timeseries of sea ice extent (top) and volume (bottom) for the annual mean (left) and the annual minima in September (right) for the individual models and the ensemble mean from CMIP5 and CMIP6 with comparison to ERA5 for the extent and PIOMAS for the*

*volume. The CMIP5 and CMIP6 models are shown in turquoise and purple respectively with a darker shade for the historical simulations and a lighter shade for the RCP8.5 and SSP585 simulations and the multi-model mean highlighted by the thick line. The blue dashed line on the sea ice extent plot marks the limit of 1 million km2, below which the Arctic is considered to be essentially ice-free.*

A big difference with the CMIP6 protocol compared to that from CMIP5 was the construction of projections for the 21$_{st}$ century (O'Neill et al., 2016). In CMIP6 these scenarios are the result of Integrated Assessment Models which aim to provide more realistic future emissions scenarios. In Figure 12 we compare projections from two of these scenarios, SSP126 and SSP585. These scenarios differ in how they assume that economic growth is fueled. In SSP126 the models assume that there is a relatively rapid uptake of non-fossil fuel-based energy sources and other sustainability measures; whereas in SSP585 the models assume that economic growth continues to be largely enabled by the use of fossil fuels. Consequently, SSP585 is a high-emissions scenario and SSP126 is relatively low-emissions scenario. Figure 12 shows the projected sea ice extent and volume from CMIP6 models under these two scenarios.

There are substantial differences between the sea ice extent and cover under the different scenarios. Under SSP585 the Arctic is projected to be nearly ice-free in September by the mid 2050s, whereas under SSP126 this isn't projected to happen. Under the SSP126 scenario the sea ice extent is projected to stabilize at around 9 million km$_2$ in the annual mean and 2.5 million km$_2$ in September by around 2060. This is in stark contrast to the SSP585 scenario where the extent continues to decline for the whole period.

There is a similar pattern with the sea ice volume. In the SSP126 scenario the volume continues to decline until the mid-21$_{st}$ century at which point it stabilizes at around 8000 km$_3$ in the annual mean and there is no significant trend in the volume for the second half of the 21$_{st}$ century. In contrast, in the SSP585 scenario the volume continues to decline for the whole of the 21$_{st}$ century. However, while the sea ice appears to stabilize by the mid-21$_{st}$ century under SSP126, in this new climatological equilibrium the sea ice extent is almost one third less than what it was in the 20$_{th}$ century equilibrium. This difference is more pronounced with the sea ice volume which stabilizes at an equilibrium two thirds lower than that of the 20$_{th}$ century.

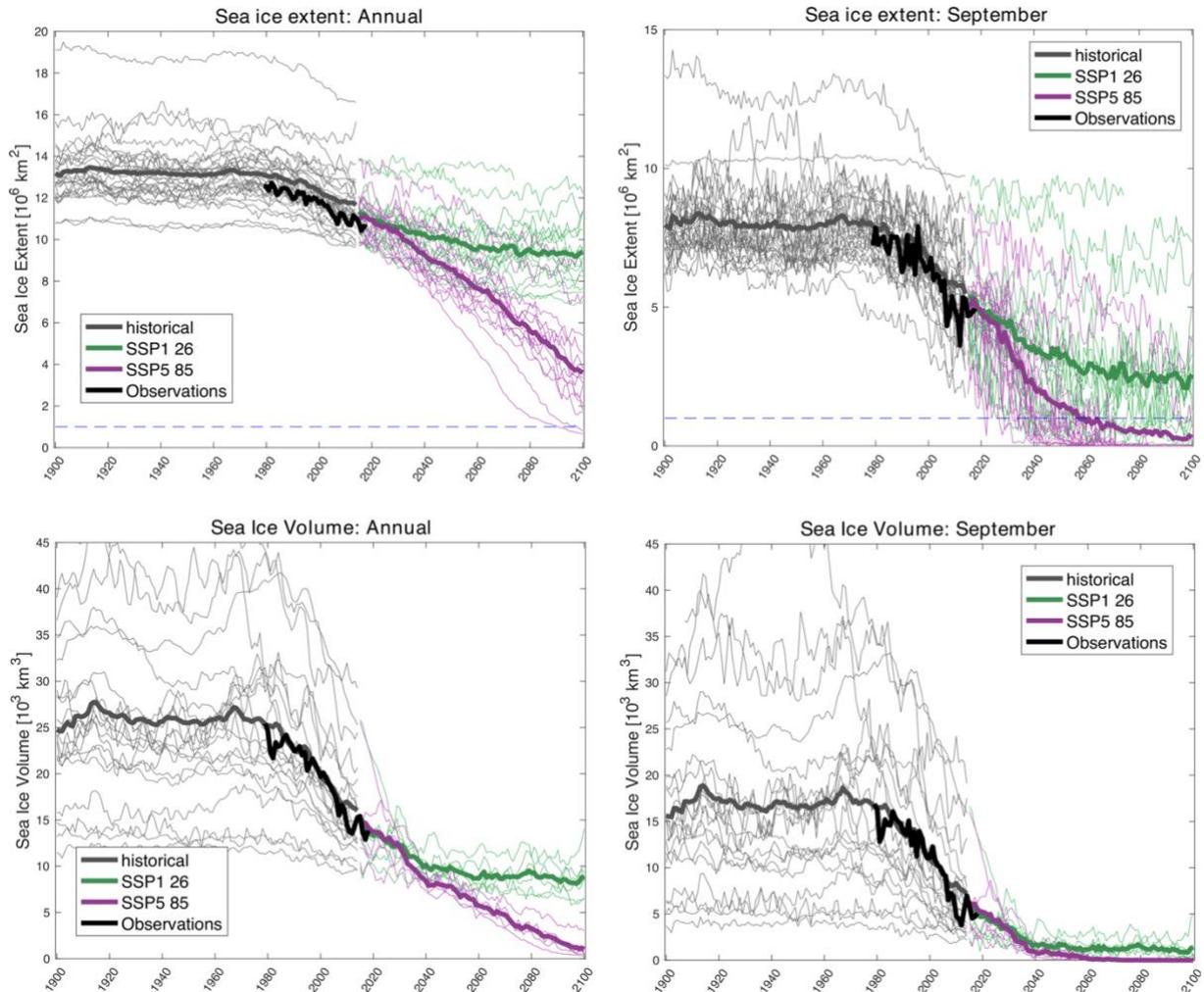

*Figure 12. Timeseries of sea ice extent (top) and volume (bottom) for the annual mean (left) and the annual minima in September (right) for the individual models and the ensemble mean from the historical (grey), SSP126 (green), and SSP585 (purple) scenarios of CMIP6 with comparison to ERA5 for the extent and PIOMAS for the volume (black). The multi-model mean highlighted by the thick lines.*

Surface air temperature

This rapid loss of sea ice is both caused by and contributes to a rapid increase in surface air temperatures in the Arctic, a phenomenon known as Arctic amplification (Serreze et al., 2007; Screen and Simmonds, 2010). Previous analysis of CMIP5 models has shown that local thermal feedbacks are the dominant driver of Arctic amplification, but that the loss of sea ice also makes an important contribution to this phenomenon (Pithan and Mauritsen, 2012).

Figure 13 shows the Arctic surface air temperature over the period of 1900-2100 contrasting the RCP8.5 scenario from CMIP5 and the SSP585 scenario from CMIP6. There is a clear, significant difference between the two scenarios for the 21st century. The multi-model means of the CMIP5 and CMIP6 historical temperatures up until 2005 are almost identical. However, the rate of warming in the CMIP6 models under the SSP585 scenario forcing is far faster than that found in the CMIP5 models under the RCP 8.5 scenario. The multi-model means of both scenarios increase almost linearly throughout the 21st century, but the faster rate of warming under SSP585 leads to an Arctic approximately 5 K warmer in 2100 than is found under RCP8.5. Note that despite the significant difference between the two results there are outliers in both model ensembles such that there is a CMIP5 model that is warmer than the CMIP6 multi-model mean and a CMIP6 model which is colder than the CMIP5 multi-model mean throughout the 21st century.

Figure 13 also contrasts the SSP126 and SSP585 scenarios of CMIP6. We can clearly see that the two scenarios are virtually identical up until around 2040. This is due to the time it takes to implement decarbonization measures and the inertia of the climate system itself. Under SSP126 we can see the Arctic temperatures stabilizing by the mid 21st century, as we saw for the sea ice (Figure 13). This new, stable Arctic climate is on average 4.7 K warmer than was found in the previous equilibrium of the early 20th century. This is almost three times the global average change in surface climate of 1.7 K under the SSP126 scenario. It is worth noting that in a recent extreme-warm year we already saw annual-mean temperatures of similar magnitude to the projected 21st century equilibrium under SSP126. This is due to the extremely high inter-annual variability found in the Arctic.

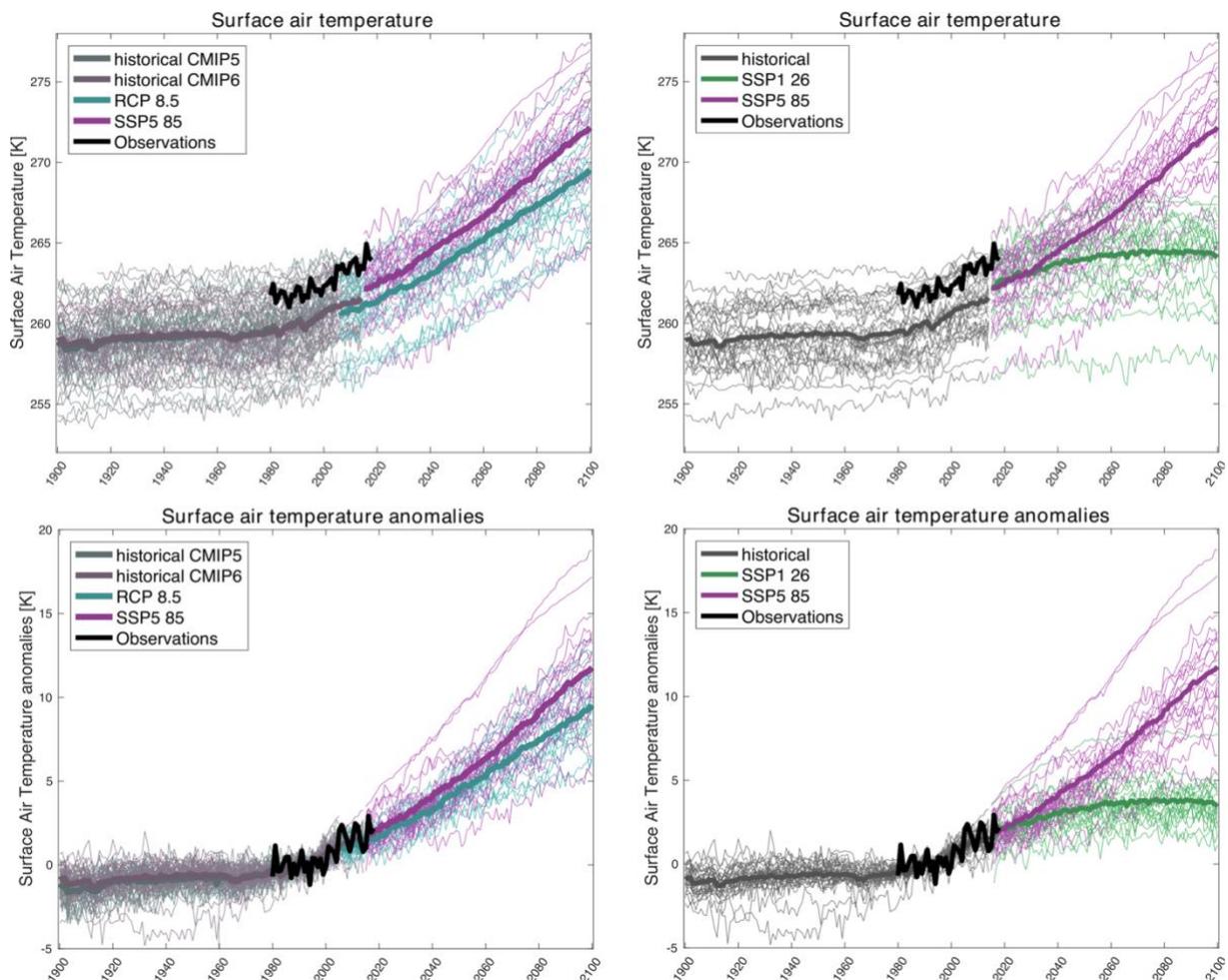

*Figure 13. Time series of the absolute and anomalies of surface air temperature over the period 1900-2100 for CMIP5 combining the historical and RCP 8.5 scenarios and CMIP6 combining the historical and SSP585 scenarios (left) and for the CMIP6 comparing the projections for SSP126 and SSP585 (right). Anomalies are taken with reference to the period 1980-2000.*

## 5. Summary

There have clearly been some improvements in CMIP6 compared to CMIP5 in capturing the state of the Arctic in the recent decades. Some key points from the inter-comparison of CMIP5 and CMIP6 skill are:

- There is better representation of the sea ice extent, edge and retreat, in particular in the Barents Sea. This was an important bias in the Arctic climate in the CMIP5 simulations, and there is a clear indication that the CMIP6 models have a much smaller bias in the mean extent, and a closer fit to the observed decline in sea ice extent in this region. This is a key location for atmospheric-ocean coupling and an improved representation of the sea ice extent here could have consequences for mid-latitude climate (Outten et al., 2013) and the representation of multi-decadal variability (Outten and Esau, 2017).
- The climatology and trend in the sea ice volume in CMIP6 shows a far better fit to PIOMAS than did CMIP5. The CMIP6 ensemble mean is an extremely good fit to the observed decline in sea ice volume over the period 1979 – 2018. However, there remains a large spread within the CMIP6 ensemble as to the climatology of the sea ice volume in the $20_{th}$ century.
- Both CMIP5 and CMIP6 models generally have too-high persistency of anomalies in the sea ice extent and volume, compared to observations. This means that these models all tend to overestimate the predictability of sea ice.
- Both the CMIP5 and CMIP6 models have on average a 4 K cold bias over the Arctic in winter (January and February). This cold bias covers most of the Arctic but is particularly strong in the regions where the models over-estimate the ice cover (the Barents Sea and the Fram Strait) and where the ice is thick (north of Greenland) or poorly resolved due to the presence of many islands not resolved in the climate models (Canadian archipelago).
- The CMIP5 and CMIP6 models tend to over-estimate the inter-annual variability in wintertime temperatures in the Arctic. This bias is larger in the CMIP6 ensemble than in the CMIP5 ensemble.
- Both the CMIP5 and CMIP6 models show a reduction in the day-to-day variability of surface air temperature. But this is not consistent with the ERA5 reanalysis which shows an increase in day-to-day variability over sea ice in winter. This could be due to missing or poorly represented surface coupling processes that are important in the winter.
- The Arctic is projected to have nearly ice-free summers by the late $21_{st}$ century in both the CMIP5 RCP8.5 and CMIP6 SSP585 scenarios. The rate of decline in sea ice extent in the period 2000-2040 is faster in CMIP6 than in CMIP5.
- In the high-emission scenario, SSP585, the Arctic is projected to be nearly ice-free in September by the mid $21_{st}$ century. However, in the low-emission SSP126 scenario the sea ice extent is projected to stabilize by around 2040 at approximately 2.5 million $km_2$. So, if the global economy rapidly decarbonizes following the SSP126 pathway there is a better than 50% chance that there will not be ice-free Septembers in the Arctic.
- The Arctic air temperatures will also stabilize by around 2040 under the SSP126 scenario at 4.7 K warmer than the 1950-1980 average which is almost three times the global average warming for this period of 1.7 K.

Acknowledgements

This publication was support by the Norwegian Research Council projects INES and KeyCLIM, and the Bjerknes Center for Climate Research's Fast Track Initiative program.

The PIOMAS reanalysis data of sea ice thickness is made available by the University of Washington (http://psc.apl.uw.edu/data/). The CMIP5 and CMIP6 data are made available by the ESGF and the data can be acquired from the ESGF nodes (e.g. https://esgf-data.dkrz.de/projects/esgf-dkrz/). The ERA5 atmospheric reanalysis data is produced and provided by the European Centre for Medium range Weather Forecasting and can be acquired via their portal (https://apps.ecmwf.int/data-catalogues/era5/).

Tables

| CMIP5 Model name | Variables available: historical | Variables available: RCP 8.5 |
|---|---|---|
| ACCESS 1.0 | SAT, PSL, SIE, SIV | SAT, PSL, SIE, SIV |
| ACCESS 1.3 | SAT, PSL, SIE, SIV | SAT, PSL, SIE, SIV |
| BCC-CSM1 | SAT, PSL, SIE, SIV | SAT, PSL, SIE, SIV |
| BCC-CSM1M | SAT, PSL, SIE, SIV | SAT, PSL, SIE, SIV |
| BNU-ESM | SAT, SIE, SIV | SAT, SIE, SIV |
| CanCM4 | SAT, SIE, SIV | SAT |
| CanESM2 | SAT, PSL, SIE, SIV | SAT, PSL, SIE, SIV |
| CCSM4 | SAT, SIV | SAT |
| CESM1-BGC | SAT, SIV | SAT |
| CESM1-CAM5 | SAT, SIV | SAT |
| CESM1-FASTCHEM | SAT, SIV | SAT |
| CMCC-CESM | SAT, SIE, SIV | SAT, SIE, SIV |
| CMCC-CM | SAT, SIE, SIV | SAT, PSL, SIE, SIV |
| CMCC-CMS | SAT, SIE, SIV | SAT, SIE, SIV |
| CNRM-CM5 | SAT, SIE, SIV | SAT, PSL, SIE, SIV |
| CNRM-CM5-2 | SIV | |
| CSIRO-Mk3.6.0 | SAT, SIE, SIV | SAT, SIE, SIV |
| CSIRO-Mk3L12 | SIE, SIV | |
| EC-EARTH | SAT, PSL, SIE, SIV | PSL, SIE, SIV |
| FGOALS-g2 | SAT, PSL, SIE, SIV | SAT, PSL, SIE, SIV |
| FGOALS-s2 | PSL, SIE, SIV | PSL, SIE, SIV |
| FIO-ESM | SAT, SIE, SIV | SAT, SIE, SIV |
| GFDL-CM2.1 | SIE, SIV | |
| GFDL-CM3 | SAT, PSL, SIE, SIV | SAT, PSL, SIE, SIV |
| GFDL-ESM2G | SAT, PSL, SIE, SIV | SAT, PSL, SIE, SIV |
| GFDL-ESM2M | SAT, PSL, SIE, SIV | SAT, PSL, SIE, SIV |
| GISS-E2-H | SAT, PSL, SIE, SIV | SAT, PSL, SIE, SIV |
| GISS-E2-H-CC | SAT, SIE, SIV | SAT, SIE, SIV |
| GISS-E2-R | SAT, PSL, SIE, SIV | SAT, PSL, SIE, SIV |
| GISS-E2-R-CC | SAT, SIE, SIV | SAT, SIE, SIV |
| HadCM3 | SAT, SIE, SIV | SAT |
| HadGEM2-AO | SIE, SIV | SIE, SIV |
| HadGEM2-CC | SAT, PSL, SIE, SIV | SAT, PSL, SIE, SIV |
| HadGEM2-ES | SAT, PSL, SIE, SIV | SAT, PSL, SIE, SIV |
| INMCM4 | SAT, PSL, SIE, SIV | SAT, PSL, SIE, SIV |
| IPSL-CM5A-LR | SAT, PSL, SIE, SIV | SAT, PSL, SIE, SIV |
| IPSL-CM5A-MR | SAT, PSL, SIE, SIV | SAT, PSL, SIE, SIV |
| IPSL-CM5B-LR | SAT, PSL, SIE, SIV | SAT, PSL, SIE, SIV |
| MIROC-ESM | SAT, PSL, SIE, SIV | SAT, PSL, SIE, SIV |
| MIROC-ESM-CHEM | SAT, PSL, SIE, SIV | SAT, PSL, SIE, SIV |
| MIROC4h | SAT, PSL, SIE, SIV | SAT, |
| MIROC5 | SAT, PSL, SIE, SIV | SAT, PSL, SIE, SIV |
| MPI-ESM-LR | SAT, PSL, SIE, SIV | SAT, PSL, SIE, SIV |
| MPI-ESM-MR | SAT, PSL, SIE, SIV | SAT, PSL, SIE, SIV |
| MPI-ESM-P | SAT, PSL, SIE, SIV | SAT |
| MRI-CGCM3 | SAT, PSL, SIE, SIV | SAT, PSL, SIE, SIV |
| MRI-ESM1 | SAT, PSL, SIE, SIV | SAT, PSL, SIE, SIV |
| NorESM1-M | SAT, PSL, SIE, SIV | SAT, PSL, SIE |
| NorESM1-ME | SAT, SIV | SAT |

Table 1. List of CMIP5 models used in the analysis presented here and which variables were available for each scenario in the analysis presented here. Variable short names are: SAT – surface air temperature; PSL – sea level pressure; SIE – sea ice extent; SIV – sea ice volume.

| CMIP6 Model name | Variables available: historical | Variables available: SSP 126 | Variables available: SSP 585 |
|---|---|---|---|
| AWI-CM1-1-MR | SAT, SIV | SAT, SIV | SAT |
| BCC-CSM2-MR | SAT, PSL, SIE, SIV | SAT, SIV | SAT, SIV |
| BCC-ESM1 | SAT, SIE, SIV | | |
| CAMS-CSM1-0 | SAT, SIE, SIV | SAT, SIE, SIV | SAT, SIE, SIV |
| CanESM5 | SAT, SIE | SAT, SIE | SAT, SIE |
| CESM2 | SAT, SIE, SIV | SAT, SIE, SIV | SAT, SIE, SIV |
| CESM2-WACCM | SAT, SIE, SIV | SAT, SIE, SIV | SAT, SIE, SIV |
| CNRM-CM6-1 | SAT, PSL, SIE, SIV | SAT, SIE, SIV | SAT, SIE, SIV |
| CNRM-CM6-1-HR | SAT, SIE, SIV | | |
| CNRM-ESM2-1 | SAT, PSL, SIE, SIV | SAT, | SAT, |
| E3SM-1-0 | SAT, SIE | | |
| EC-Earth3 | SAT, PSL, SIE, SIV | SAT, SIE | SAT, SIE |
| EC-Earth3-Veg | SAT, PSL, SIE, SIV | SAT, SIE | SAT, SIE |
| FGOALS-f3-L | SAT | SAT | SAT |
| FGOALS-g3 | SAT, PSL | SAT | SAT |
| GFDL-CM4 | SAT, PSL, SIE, SIV | | SAT, SIE, SIV |
| GFDL-ESM4 | SAT, SIE, SIV | SAT, SIE, SIV | SAT, SIE, SIV |
| GISS-E2-1-G | SAT, PSL, SIV | | |
| GISS-E2-1-G-CC | SAT, SIV | | |
| GISS-E2-1-H | SAT, SIE, SIV | | |
| HadGEM3-GC31-LL | SAT, SIE, SIV | | |
| INMCM-4.8 | SAT, SIE | SAT, SIE | SAT, SIE |
| INMCM-5.0 | SAT, SIE | | |
| IPSL-CM6A-LR | SAT, PSL, SIE, SIV | SAT, SIE, SIV | SAT, SIE, SIV |
| MCM-UA-1-0 | SAT, SIE | SAT | SAT |
| MIROC6 | SAT, SIE | SAT, SIE | SAT, SIE |
| MIROC-ES2L | SAT, SIE | SAT, SIE | SAT, SIE |
| MPI-ESM1-2-HR | SAT, PSL, SIE, SIV | SAT, SIE, SIV | SAT, SIE, SIV |
| MRI-ESM2-0 | SAT, PSL, SIE | SAT, SIE | SAT, SIE |
| NESM3 | SAT, PSL, SIE | SAT, SIE | SAT, SIE |
| NorCPM1 | SAT, SIE | | |
| NorESM2-LM | SAT, SIE, SIV | | |
| SAM0-UNICON | SAT, PSL, SIE, SIV | | |
| UKESM1-0-LL | SAT, SIE, SIV | SAT, SIE, SIV | SAT, SIE, SIV |

Table 2. List of CMIP6 models used in the analysis presented here and which variables were available for each scenario used in the analysis presented here. Variable short names are: SAT – surface air temperature; PSL – sea level pressure; SIE – sea ice extent; SIV – sea ice volume.